\documentclass[a4paper,fleqn,usenatbib]{mnras}

\usepackage{newtxtext,newtxmath}

\usepackage[T1]{fontenc}
\usepackage{ae,aecompl}

\usepackage{graphicx}	
\usepackage{amsmath}	
\usepackage{amssymb}	

\newcommand{\removedtext}[1]{}

\newcommand{\addedtext}[1]{#1}

\title[The nature of the photometric variability of $\varphi$~Phe]{The nature of
the photometric variability of HgMn stars: A test of simulated light curves of $\varphi$~Phe against the TESS data}
\author[M. Prv\' ak et al.]{
Milan Prv\' ak,$^{1}$\thanks{E-mail: prvak@physics.muni.cz}
Ji\v r\' i Krti\v cka$^{1}$
and Heidi Korhonen$^{2}$
\\
$^{1}$Department of Theoretical Physics and Astrophysics, Masaryk
              University, CZ-611\,37 Brno, Czech Republic\\
$^{2}$DARK, Niels Bohr Institute, Lyngbyvej 2, DK-2100 Copenhagen, Denmark\\
}

\date{Accepted XXX. Received YYY; in original form ZZZ}

\pubyear{2019}

\begin{document}
\label{firstpage}
\pagerange{\pageref{firstpage}--\pageref{lastpage}}
\maketitle

\begin{abstract}
    The inhomogeneous surface distribution of heavy elements is known to
    cause periodic light variability of 
    magnetic chemically peculiar stars. It is unclear to what extent the same
    paradigm applies to mercury-manganese (HgMn) stars.
    We aim to model the photometric variability of the HgMn star $\varphi$~Phe
    using abundance maps obtained from high-resolution spectroscopy and to study
    how this variability evolves with time.
    We compute a grid of {\sc atlas\addedtext{12}} model atmospheres and the corresponding {\sc synspec}
    synthetic spectra. Interpolating within this grid and integrating the
    specific intensity over the visible stellar surface at different rotational
    phases, we obtain theoretical light curves of the star.
    We predict the variability of $\varphi$~Phe 
    \addedtext{in the ultraviolet and in the visible spectral regions}
    with amplitude of the order of
    millimagnitudes,
    \addedtext{
    mainly caused by absorption in lines of yttrium, chromium, and
    titanium.
    }
    We also show how this variability is affected by changes
    of the distribution of the heavy elements over time.
    \removedtext{We show that the star $\varphi$~Phe exhibits millimagnitude
    variability in the ultraviolet and in the visible spectral regions,}
    \removedtext{
        This variability is
    mainly caused by absorption in lines of yttrium, and
    strontium.
    }
    The main characteristics of the predicted light variability of
    $\varphi$~Phe correspond \addedtext{roughly} to the variability of the star
    observed with the {\it TESS} satellite.
\end{abstract}




\section{Introduction}
Mercury-manganese (HgMn) stars, also known as CP3 stars, typically feature
an increased and inhomogeneous surface distribution of heavy elements, such as
Hg, Ga, Mn, Y, Sr, or Cr \citep{castelli2004,monier2015}.
Unlike most other types of chemically peculiar stars with abundance spots,
HgMn stars do not have strong organised magnetic fields 
\citep[e.g.,][]{hubrig2012,kochukhov2013,catanzaro2016}. Slow rotation,
atomic diffusion, and a lack of sub-surface convective zone are probably
the cause of their abnormal chemical composition \citep{michaud1976}. HgMn
stars are predominantly found in binary systems \citep{budaj1996,scholler2010A}.

Line profile variations and inhomogeneous surface distribution of elements have
been observed in some of HgMn stars
\citep[e.g.,][]{briquet2010,makaganiuk2011}. Also, evidence of the secular
evolution of the surface abundance structures has been provided 
\citep[e.g.,][]{kochukhov2007,hubrig2010}. However, observations of the photometric
variability of HgMn stars are still relatively rare \citep{paunzen2018}.

When the line profile variations are observed in a star, Doppler imaging
can be used to derive the surface distribution of chemical
elements if the rotation is fast enough~\citep[e.g.,][]{rice1989,piskunov2002,luftinger2010}. Similarly, when
abundance maps are available, we can use model atmospheres and synthetic
spectra to predict photometric variability of the star. Comparing the light
curve obtained in this way with the observed photometric variability of the
star (where available) can also serve as a verification of the abundance maps,
atomic data and our comprehension of stellar atmospheres. This method has been
used with success several times in the past \citep[e.g.,][]{krticka2009,prvak2015}.

However, this method cannot be easily applied to HgMn stars. Doppler imaging
requires relatively bright targets, while the photometry with
sufficient precision can only be derived using space-borne photometry, which
is typically possible \addedtext{only} for relatively faint stars \citep[e.g.,][] 
{hummerich2018}, making combining the two techniques quite difficult. Therefore,
only a comparison of general properties of predicted and observed light curves
is currently feasible for HgMn stars. Recently, the Transiting Exoplanet Survey
Satellite ({\it TESS}) provided us with the capability to observe relatively
bright targets with sufficient precision, making the detection of the photometric
variability possible for at least some HgMn stars \citep[see e.g.,][]{sikora2019}.

In the present work, we want to investigate the effect of the presence of
heavy elements on the spectral energy distribution (SED) of the emergent
radiation, to show whether or not it leads to photometric variability, and 
to study how this variability is affected by the secular evolution of the 
surface abundance spots. Preliminary results were published in conference
proceedings \citet{prvak2018}. Our object of interest is a HgMn star $\varphi$~Phe
(HR 558, HD 11753).

The star $\varphi$~Phe is a spectroscopic binary~\citep{pourbaix2013}. The inhomogeneous
distribution of heavy elements was investigated by~\citet{briquet2010} and
later revisited in~\citet{korhonen2013}. \addedtext{Only weak magnetic fields or
no fields have been reported in $\varphi$ Phe \citep{makaganiuk2012,hubrig2010}.}
    \begin{table}
      \caption[]{Parameters of the star $\varphi$~Phe
      \citep[after][]{korhonen2013}.}
         \label{TabParam}
         \centering
         \begin{tabular}{l r}
            \hline
            Effective temperature $T_\mathrm{eff}$    & 10600\,K           \\
            Surface gravity $\log g$ [cgs]            & 3.79               \\
            Rotational period $P$                     & 9.53077\,d           \\
            Rotational velocity projection $v \sin i$ & 13.5\,km\,s$^{-1}$ \\
            Inclination $i$                           & 53$^\circ$         \\
            \hline
         \end{tabular}
   \end{table}
   \begin{figure}
   \centering
   \includegraphics[width=8cm]{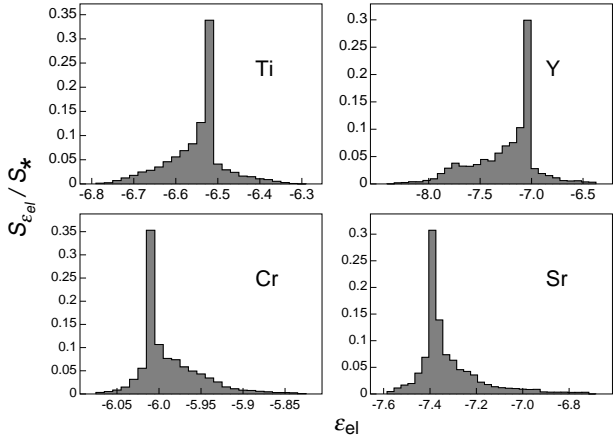}
      \caption{Distribution of heavy element abundances on the surface of 
               $\varphi$~Phe. \addedtext{The values are averaged for the four epochs.}}
      \label{FigHist}
   \end{figure}
        \begin{table}
            \caption{Abundances used for the calculation of grid of model
            atmospheres and synthetic spectra. \addedtext{Solar abundance \citep[see e.g.,][]{grevesse1998} was adopted for the other elements.}}
            \label{tab-abun}
            \begin{center}
                \begin{tabular}{llllll}
                    \hline
                    Element & \multicolumn{5}{l}{Abundances
                    $\varepsilon_\mathrm{el}$
                    }                \\
                    \hline
                    Cr      &  -6.1   &  -5.8   &         &         &       \\
                    Ti      &  -6.8   &  -6.3   &         &         &       \\
                    Sr      &  -7.6   &  -7.1   &  -6.6   &         &       \\
                    Y       &  -8.3   &  -7.8   &  -7.3   &  -6.8   & -6.3  \\
                    \hline
                \end{tabular}
            \end{center}
        \end{table}
   \begin{figure*}
   \centering
   \includegraphics[width=15cm]{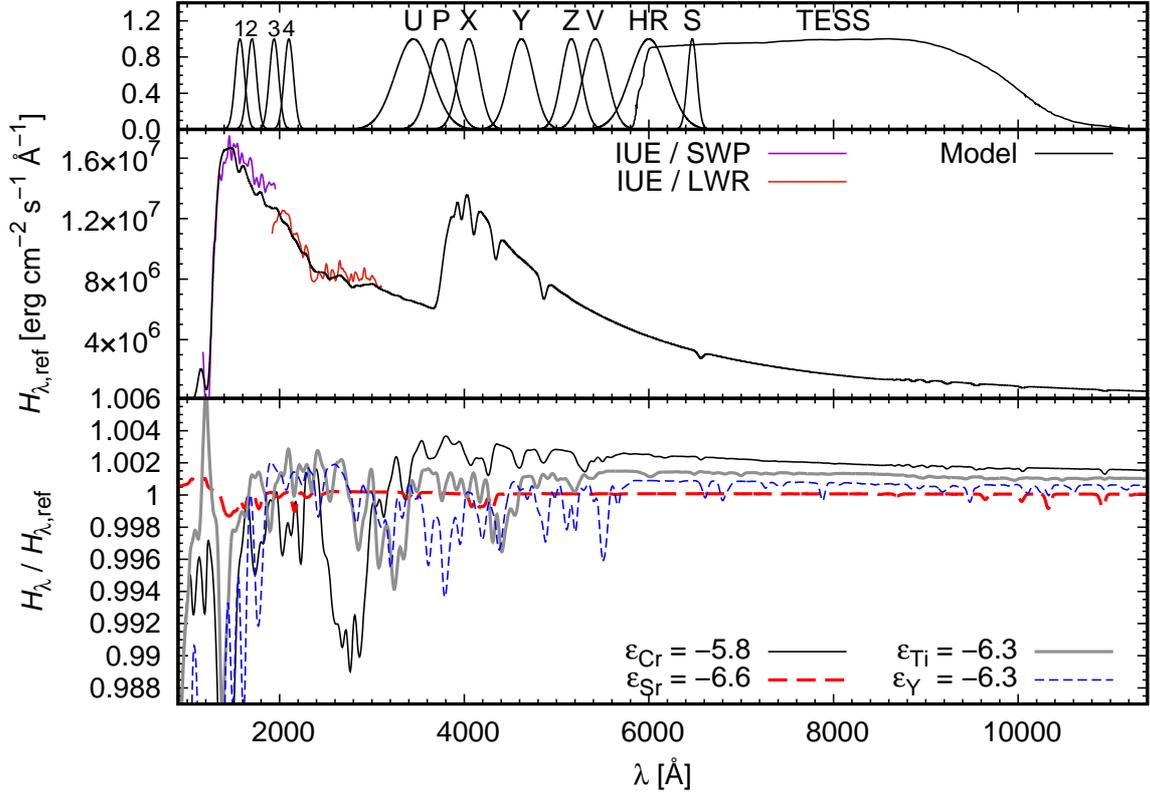}
   \caption{
       \addedtext{
           {\em Upper panel:} Response functions of the passbands used in the paper, namely the artificial
           gaussian passbands centered at 1570\,\AA, 1700\,\AA, 1940\,\AA, and 2100\,\AA, labeled 1, 2, 3, and 4, respectively,
           the gaussian functions used to approximate the $U$, $P$, $X$, $Y$, $Z$, $V$, $HR$, and $S$ passbands of
           the ten-color photometric system, and the {\it TESS} passband based on \citet{ricker2015}. The response
           functions are scaled to have a maximum value of unity.
           {\em Middle panel:} Emergent flux from a reference model atmosphere
           with $\varepsilon_\mathrm{Cr} = -6.1$, $\varepsilon_\mathrm{Ti} = -6.8$,
           $\varepsilon_\mathrm{Sr} = -7.6$, and $\varepsilon_\mathrm{Y} = -8.3$, compared
           with spectra obtained using the {\it SWP} and {\it LWR} intruments of the International
           Ultraviolet Explorer ({\it IUE}).
           {\em Lower panel:}
           Emergent flux from model atmospheres with enhanced
           abundance of the individual elements minus the flux from the reference
           model. The model fluxes were smoothed with Gaussian filters (standard deviation $\sigma = 25$\,\AA{ }
           for the models and 10\,\AA{ }for the {\it IUE} spectra) to emphasize the changes in continuum.
       } 
   } 
         \label{FigSED}
   \end{figure*}
   \begin{figure}
   \centering
   \includegraphics[width=7cm]{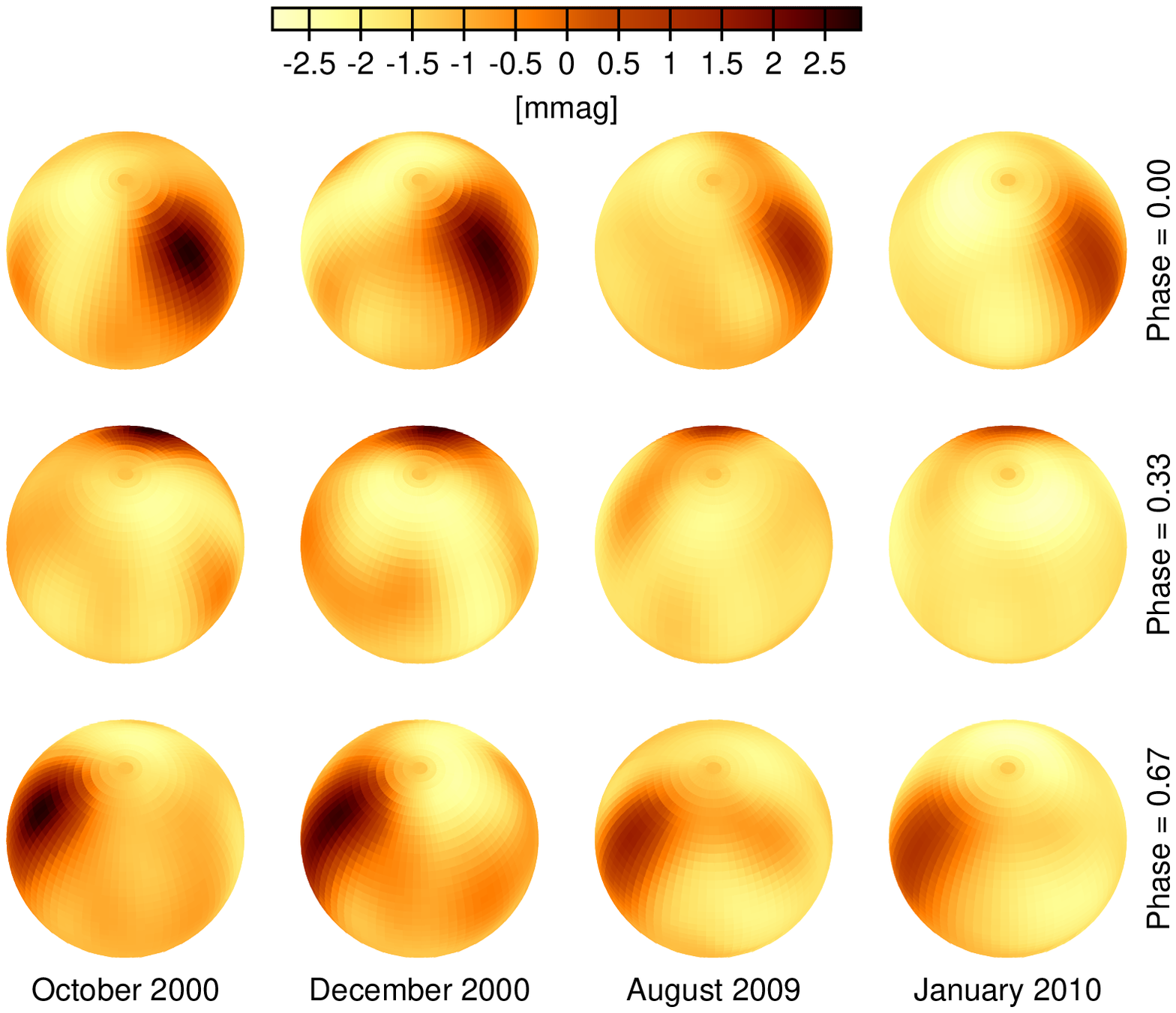}
   \includegraphics[width=7cm]{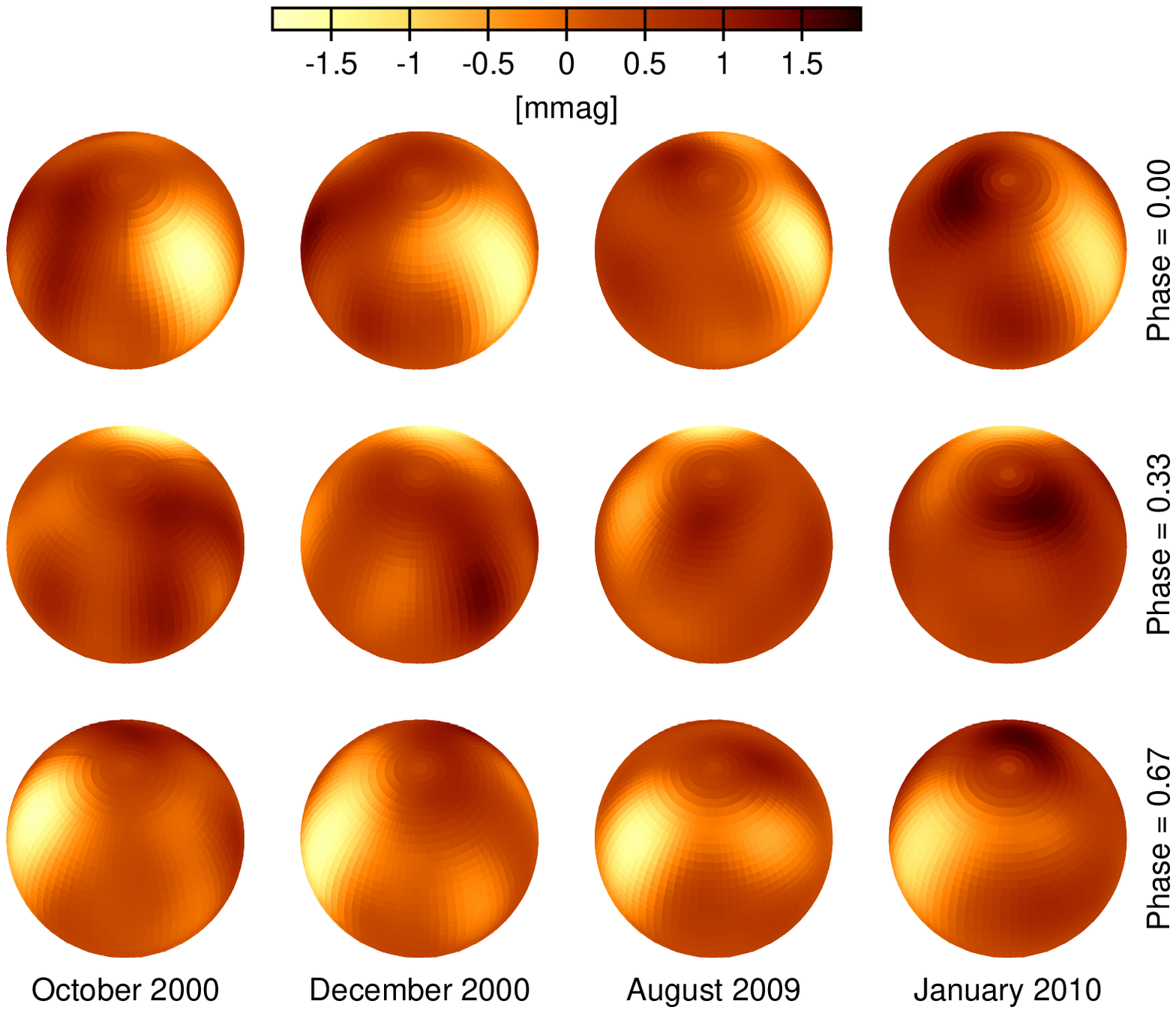}
      \caption{
          \addedtext{
            Time evolution of the emergent flux from the surface of the star
            at different rotational phases in a gaussian passband
            centered at 1700~\AA{ }with a standard deviation of 50\,\AA{ }({\em upper panel})
            and the {\it TESS} passband ({\em lower panel}).
          }
      }
      \label{FigSpheres}
   \end{figure}
   \begin{figure*}
   \centering
   \includegraphics[width=13cm]{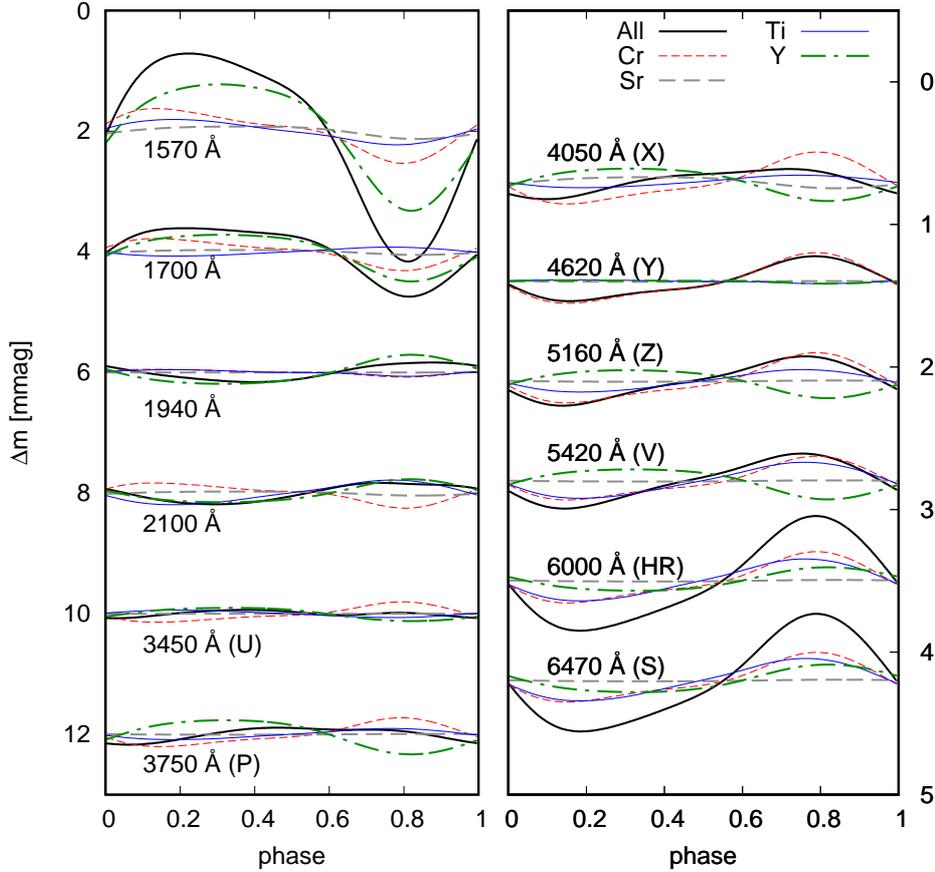}
   \caption{
       \addedtext{
         Light curves of $\varphi$~Phe computed using only the abundance
         maps of the individual elements (January 2010), along with the overall
         variability of the star (solid line) in different passbands. The light
         curves have been shifted vertically to better demonstrate the 
         variability and marked by central wavelengths of the corresponding filters.
         Please note the different vertical scale of the two panels.
       }
   }
   \label{FigComp}
   \end{figure*}
   \begin{figure}
   \centering
   \includegraphics[width=8cm]{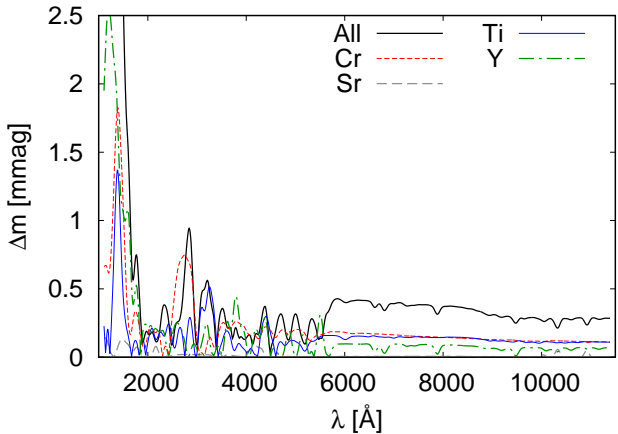}
   \caption{
     \addedtext{
         Amplitude of the photometric variability of $\varphi$~Phe (January 2010) computed
       as a function of wavelength in Gaussian passbands with a 
       standard deviation of 50~\AA.
     }
   }
   \label{FigAmp}
   \end{figure}
   \begin{figure*}
   \centering
   \includegraphics[width=13cm]{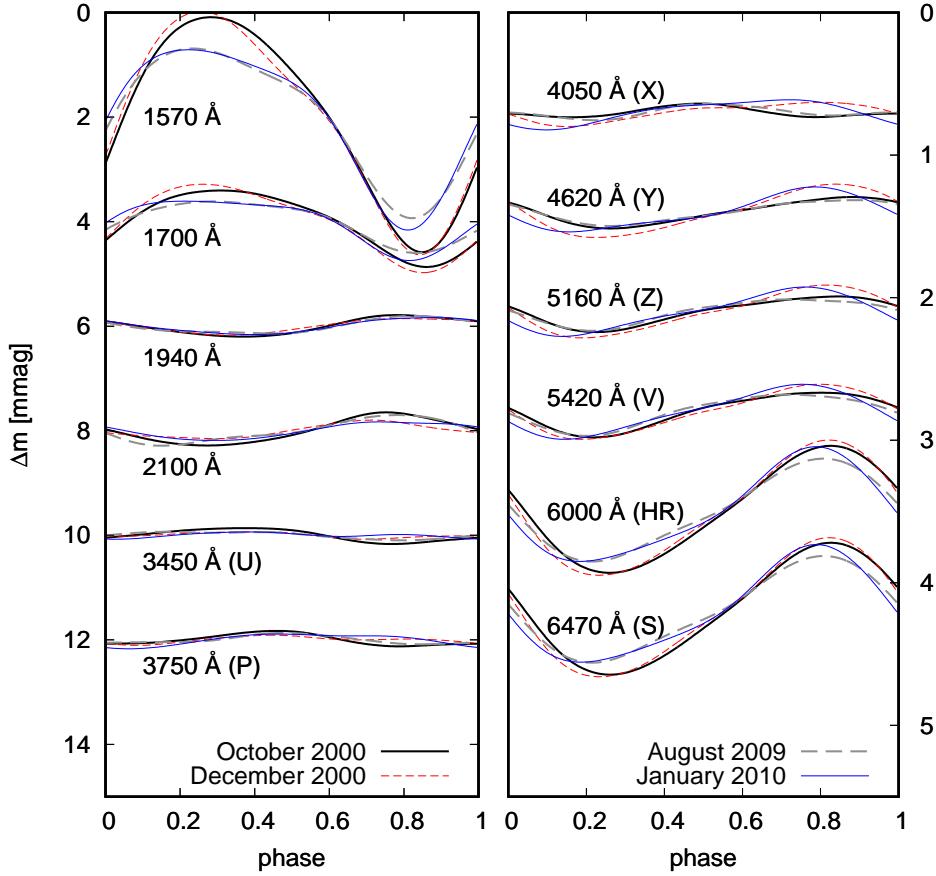}
      \caption{The light variability of $\varphi$~Phe derived from abundance
         \addedtext{
          maps obtained at different times. The light curves have been shifted
          vertically to better demonstrate the variability and marked 
          by central wavelengths of the corresponding filters.
          Please note the different vertical scale of the two panels.
        }
      }
         \label{FigEvo}
   \end{figure*}
   \begin{figure}
   \centering
   \includegraphics[width=8cm]{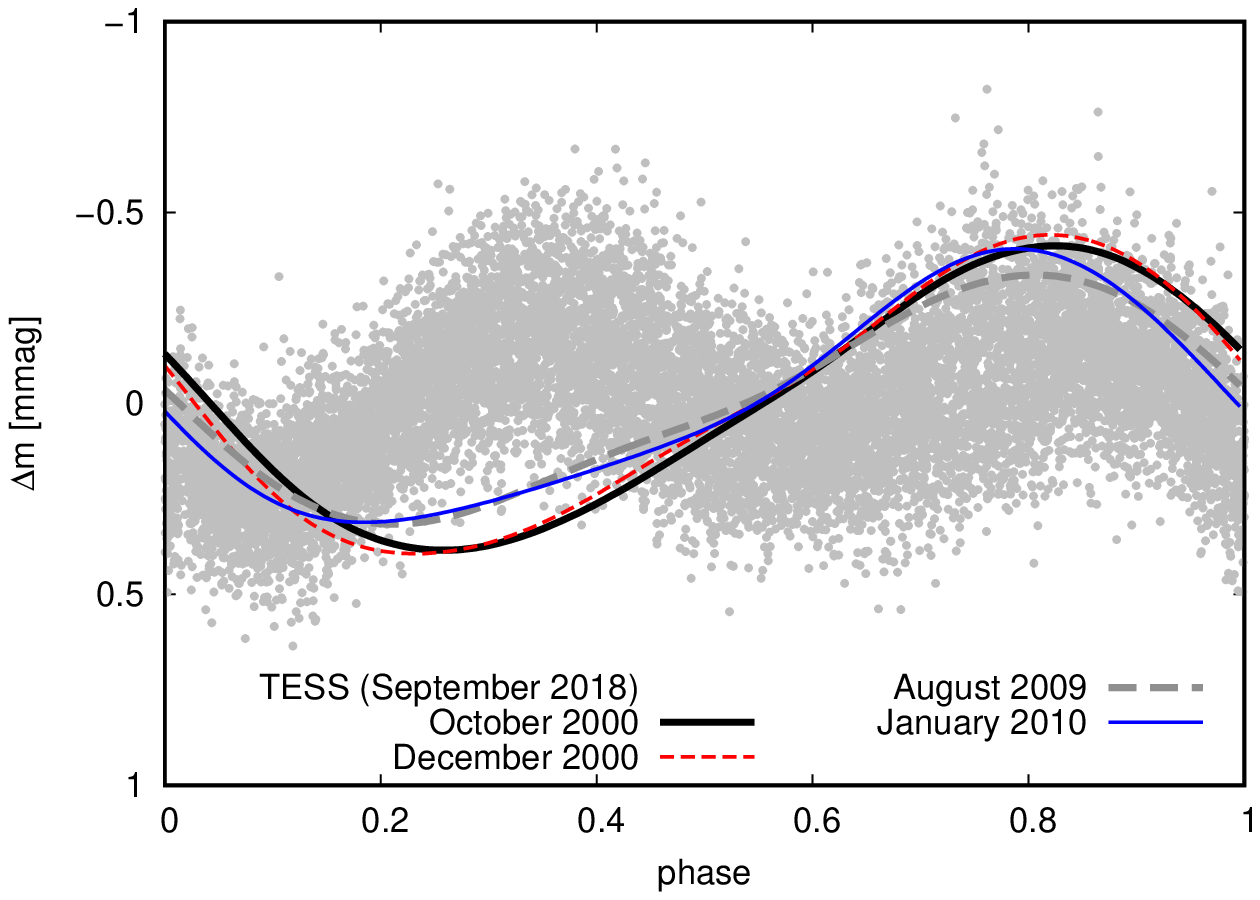}
      \caption{
        \addedtext{
          The light variability of $\varphi$~Phe derived from abundance
          maps obtained at different times compared with the phtometric variability
          of the star observed by {\it TESS}.
        }
      }
         \label{FigObs}
   \end{figure}

\section{Methods}
We used a series of abundance maps obtained using the {\it CORALIE} spectrograph
in October 2000, December 2000, August 2009 and January
2010~\citep{korhonen2013}. The overall distribution of chromium, titanium,
strontium, and yttrium is shown as histograms in Fig.~\ref{FigHist}. Here, the
abundance of each element is expressed as $\varepsilon_{\mathrm{el}} = 
\log \frac{N_{\mathrm{el}}}{N_{\mathrm{H}}}$. The vertical axis of 
Fig.~\ref{FigHist} shows the total area $S_{\varepsilon_\mathrm{el}}$, where the
abundance of the given element lies within the interval 
$(\varepsilon_\mathrm{el} - \delta / 2; \varepsilon_\mathrm{el} + \delta / 2)$,
where $\delta$ corresponds to the width of a column in the histogram.
In Fig.~\ref{FigHist}, $S{\!}_\ast$ denotes stellar surface area.

We applied the {\sc atlas\addedtext{12}} code \citep{kurucz1996,castelli2005}
to compute a grid of model atmospheres covering the relevant range of
abundances in the maps.
The physical parameters of the star are listed in
Table~\ref{TabParam}. Table~\ref{tab-abun} shows the abundance values used
for each individual element of the abundance grid.
The atomic data were taken from the Kurucz's
website\footnote{http://kurucz.harward.edu}. We applied the {\sc synspec} code
(with atomic data from \citealt{lanz2007}) to
produce the corresponding synthetic spectra for each model atmosphere. By means
of multilinear interpolation within the grid of synthetic spectra and integrating
over wavelength, we calculated corresponding angle-dependent specific intensities
for each of the 240 $\times$ 120 elements of the stellar surface. Summing over the
visible surface of the star, we obtained the emergent flux and, consequently,
the light curves of the star (see~\citet{prvak2015} for more details).
\addedtext{We adopted the ephemeris $JD = 2\,451\,800.0 + 9.53077$ from \citet{korhonen2013}.}

\section{Results}
\subsection{Influence of chemical elements}
The presence of heavy elements in the photosphere causes some portion of the
spectral energy to be absorbed mainly by bound-bound transitions in the
ultraviolet region of the spectrum. This energy is then re-emitted mostly in
the near-ultraviolet and in the visible. The effect this process has on the
spectral energy distribution (SED) is shown in Fig.~\ref{FigSED}. Because the
absorption occurs mostly in lines, while the re-emission contributes mostly in
continuum, the resulting effect varies very strongly with wavelength.

As a result, the parts of the stellar surface with higher concentration of
heavy elements will appear as dark spots in the far-UV, while the same spots
will show as bright in the visible. The total flux integrated over the entire
spectrum, though, is conserved. Fig.~\ref{FigSpheres} shows the emergent flux
from the surface of the star in two different passbands, \removedtext{centered at 1300\,\AA
{} and 4050\,\AA, respectively.} \addedtext{a gaussian function centered at 1700\,\AA{ }and the 
red--infrared {\it TESS} passband (see Fig.~\ref{FigSED} for the response functions of the various
passbands used in this paper). } Because there is more than one element
contributing to the spectral energy redistribution, and the absorption lines of
these individual elements dominate different regions of the spectrum, the shape
of the bright and dark spots on the stellar surface varies with wavelength.

Predicted flux distribution for reference abundances agrees reasonably with the {\it IUE}
observations ({\it SWP} 19761 and {\it LWR} 15774)\footnote{Data downloaded from the Mikulski
Archive for Space Telescopes (MAST) at https://archive.stsci.edu/iue/}
in Fig.~\ref{FigSED}. Small differences are most likely caused by missing opacity in model atmospheres.

\subsection{Effect of the individual elements}

The inhomogeneous distribution of flux in most spectral bands, together with
stellar rotation, leads to periodic photometric variability of the star.
Fig.~\ref{FigComp} shows the overall modelled photometric variability of the
star, together with the photometric variability computed only taking into
account the effect of each individual chemical element separately, to show the
contribution of each element to the overall variability of the star. The central
wavelengths and \removedtext{dispersion}\addedtext{half-widths} of the filters used in the plots are taken from
\citet{prvak2015}. While the shape of the curves of the individual elements
remains roughly the same at all wavelengths, being ``composed'' in different
ratios results in a curve of varying shape.

Fig.~\ref{FigAmp} shows the dependence of the amplitude of the photometric
variability on wavelength. Again, the amplitude of the overall variability, defined as
one-half of the difference between the maximum and the minimum magnitude of the star, is
plotted together with the amplitude of the variability computed only taking into
account the influence of the four chemical elements, one at a time. 
\removedtext{
The variability of the star is largest in the visible and the near-UV parts of
the spectrum, especially at wavelengths around 3000--4000~\AA.}
\addedtext{The variability of the star is largest in the far-UV part of the spectrum, especially around 
$1300$--$1400$\,\AA, mainly because of the very low overall flux in this region. In the near-UV and the optical
region up to about $6000$\,\AA, the amplitude in narrow passbands varies significantly with wavelength due to the significant
absorption in lines. In the red and the infrared, the variability is mostly governed by re-emission in continuum and the
amplitude is relatively constant throughout the region.}

Yttrium, which shows the most pronounced abundance variation, seems to also have the
strongest effect on the variability, especially in the visible part of the
spectrum. The effect of titanium \addedtext{and chromium} is also significant. \removedtext{Chromium and} Strontium
only contributes to the overall variability marginally.

Fig.~\ref{FigEvo} compares the overall modelled photometric variability of the
star in four different times corresponding to the four sets of the abundance
maps available. The secular evolution of the abundance spots causes a small but
visible change in the shape of the light curves. The changes in the shape of
the light curves are mostly caused by the secular evolution of spots of 
titanium and, to a lesser degree, of chromium and yttrium.

There are space-borne photometric observations of the star obtained by {\it TESS}
\citep{ricker2015},
spanning a period of 30 days in September 2018, of which 21 days of usable data
are left after reduction and detrending. This photometry is shown in 
Fig.~\ref{FigObs} together with our synthetic light curves corresponding to the
four individual sets of abundance maps. \removedtext{The passband used by {\it TESS}, spanning
into the infrared part of the spectrum, is broader than the $S$ passband of the
ten-colour photometric system used for the light curves shown in the plot.}
Unfortunately, there are no abundance maps available for the epoch corresponding to the {\it TESS}
observations. For \removedtext{these reasons}\addedtext{this reason}, the comparison is only
qualitative. 
\removedtext{The amplitude of the observed light variability is smaller than the models indicate, but the difference
is comparable to the difference in amplitude between light curves predicted for individual epochs.}
\addedtext{The amplitude of the predicted light curve agrees reasonably well with the observations. However,
the double-wave form of the observed variability is not reproduced, the brightness increase with maximum at
approximately the phase 0.35 is missing.}

\section{Discussion}

Our results about the variability of the studied HgMn star nicely agree with
general characteristics of variability of this type of stars derived using space
borne instruments. For example, \citet{hummerich2018} detected periodical light
variations in HgMn star KIC 6128830 with amplitude 3.4~mmag and period
about 5~days. These variations are fully comparable to the light variations of
$\varphi$~Phe. Also HgMn star HD 45975 shows a similar time-scale and
amplitude of variability as $\varphi$~Phe \citep{morel2014}. The abundance
variations in some HgMn stars are likely so weak, that the corresponding light
variability evades detection even with modern instruments
\citep{ghazaryan2013}. On the other hand, larger surface overabundance is
required to explain the rotational light variability of HgMn stars with
amplitudes of about 0.01~mag \citep{strassmeier2017}.

The predicted variability of the star does not agree completely with the observations
by {\it TESS}. 
\removedtext{However, }The discrepancy between the models and the observations 
\removedtext{is comparable to the differences between the light curves corresponding to the individual sets of
abundance maps and} can \removedtext{therefore} be partially explained by the ongoing secular evolution of
the surface abundance structures between the years 2010 and 2018.
The changes in the abundance maps are rather small, so one would not really expect drastic changes
like seen between the {\it TESS} light-curve and the light-curves obtained from the Doppler images.
\removedtext{Apparently, the high abundance spot at phase 0.8 gets markedly weaker with time, as is also reflected by 
the predicted photometry. Possibly, the spot has gotten even weaker and now only results to the amplitude
seen in the {\it TESS} data. Similarly, a new spot could have formed at around phase 0.3 that explains
the double humped light-curve.}
\addedtext{The discrepancy is strongest at phases about 0.2--0.4. However, this is also the part of the light curve, where
the shape of the light curve changes most drastically between the individual datasets. Possibly, a new spot could have
formed here, eventually leading to the double humped light curve.} Obtaining new spectroscopic observations of the
star to verify whether or not this is the case would be very useful.

\removedtext{Also, the photometric
passband used for the light curve synthesis does not perfectly match the {\it TESS} passband,
which is significantly broader and situated farther towards the infrared region, which
could have caused a difference in the resulting amplitude of the light variability.
On the other hand, the shapes of the light curves do not significantly vary with
wavelength in the red part of the spectrum, which makes this assumuption justifiable.}

Additionally, abundance structures of chemical elements other than the four considered in our models could
contribute to the light variability, \removedtext{reducing the amplitude} \addedtext{changing the amplitude or shape
of the light curve}.
Missing opacity due to additional elements could explain small differences
between predicted and observed fluxes in Fig.~\ref{FigSED}.
Finally, the binarity of the star $\varphi$~Phe may have affected the amplitude of both the observed and
the modelled light variability.
\addedtext{However, \citet{pourbaix2013} reports a magnitude difference of 5.7 in $V$ and 3.9 in $K$. It is 
therefore unlikely that the secondary component could have affected the light variability significantly.}

Our models also indicate that the secular evolution of the abundance structures causes a gradual
shift in the phases of photometric minima and maxima. That may affect adversely the determination of
the rotational period of the star, which has never been established with certainty
\citep[see, e.g.,][]{korhonen2013}. An incorrectly determined period, on the other hand, may make a
false impression of the evolution of the abundance structures.

Contrary to many other chemically peculiar stars, the mechanism of the light
variability of $\varphi$~Phe cannot be tested in the UV with the available instruments.
The amplitude of the flux variability in the UV is only a few millimagnitudes (see
Fig.~\ref{FigComp}) and is therefore an order of magnitude smaller than the
precission of the typical flux calibrated UV spectrographs, which is about a few
percentage points \citep{stis}.

Our study of the light variability of $\varphi$~Phe may provide a clue for the
explanation of periodical low-amplitude light variability in A type stars
derived using Kepler satellite \citep{balona2011,balona2013}. The amplitude of
the detected light variability is similar to that predicted here. Such
variations can be tested using detailed spectroscopy. 

We predicted photometric variability of $\varphi$~Phe. However, there are some
things to keep in mind. First of all, the models used for our calculations \removedtext{are}
all \addedtext{assume} LTE. This may be less precise than using NLTE models. However, as shown
by~\cite{krticka2012}, NLTE effects do not affect the SED
variability significantly.

Also, the Doppler imaging technique uses a single model atmosphere,
corresponding to the mean chemical composition for the given star. While
varying the chemical abundances during the mapping process, these modified
values are used to produce changes in the modelled line profiles. However, the
model atmosphere is not changed accordingly. This means that the effect the
modified chemical composition has on the physical structure of the atmosphere
is not taken into account. This may also affect the accuracy of our result.
The acceptability of this simplification has been questioned by 
\citet{stift2012}, but \citet{kochukhov2012} showed that this approximation
does not significantly influence the result, with the exception of cases with
extreme overabundance of heavy elements. The good agreement of our previous
works with observations supports this conclusion.

\section{Conclusions}

We analysed the photometric variability of $\varphi$~Phe. Using the abundance
maps of the star, we predicted the star exhibits photometric
variability in the ultraviolet and the visible parts of the spectrum.
\removedtext{The amplitude of the variability is greatest in the range of
wavelengths between 3000 and 4000~\AA, where it reaches almost 2 
millimagnitudes.} The amplitude of the variability is approximately 0.5 millimagnitudes
in most parts of the near-UV and the optical spectrum. This is a relatively small variability
and it is difficult to verify experimentally. However, we were able to discern the variability
in the {\it TESS} photometry. Unfortunately, due to the large time gap between the
sets of abundance maps and the photometric observations \addedtext{and possibly missing information
on the abundance distribution of the other chemical elements}, the comparison we made
is only qualitative. 

The variability of the star is mainly caused by bound--bound
transitions of yttrium, \addedtext{chromium}, and titanium. The increased abundance of heavy
chemical elements such as yttrium is common in HgMn stars. The research of these
stars, of their atmospheres and variability would certainly benefit from 
availability of more detailed, complete and accurate atomic data.

We also show that the secular evolution of the spots, especially those of
titanium and chromium, are reflected in the changes of the photometric
variability of the star.

Our results contribute to the rather small amount of cases of the observed
photometric variability in HgMn stars. Hopefully, it will also provide verification
of the abundance maps of the star and the Doppler imaging technique in general,
the atomic data and the theory of stellar atmospheres.

\section*{Acknowledgements}

MP and JK were supported by grant GA \v CR
18-05665S.
HK acknowledges support from the Augustinus foundation.




\bibliographystyle{mnras}
\bibliography{phiphe} 

\begin{thebibliography}{}
\makeatletter
\relax
\def\mn@urlcharsother{\let\do\@makeother \do\$\do\&\do\#\do\^\do\_\do\%\do\~}
\def\mn@doi{\begingroup\mn@urlcharsother \@ifnextchar [ {\mn@doi@}
  {\mn@doi@[]}}
\def\mn@doi@[#1]#2{\def\@tempa{#1}\ifx\@tempa\@empty \href
  {http://dx.doi.org/#2} {doi:#2}\else \href {http://dx.doi.org/#2} {#1}\fi
  \endgroup}
\def\mn@eprint#1#2{\mn@eprint@#1:#2::\@nil}
\def\mn@eprint@arXiv#1{\href {http://arxiv.org/abs/#1} {{\tt arXiv:#1}}}
\def\mn@eprint@dblp#1{\href {http://dblp.uni-trier.de/rec/bibtex/#1.xml}
  {dblp:#1}}
\def\mn@eprint@#1:#2:#3:#4\@nil{\def\@tempa {#1}\def\@tempb {#2}\def\@tempc
  {#3}\ifx \@tempc \@empty \let \@tempc \@tempb \let \@tempb \@tempa \fi \ifx
  \@tempb \@empty \def\@tempb {arXiv}\fi \@ifundefined
  {mn@eprint@\@tempb}{\@tempb:\@tempc}{\expandafter \expandafter \csname
  mn@eprint@\@tempb\endcsname \expandafter{\@tempc}}}

\bibitem[\protect\citeauthoryear{{Balona}}{{Balona}}{2011}]{balona2011}
{Balona} L.~A.,  2011, \mn@doi [\mnras] {10.1111/j.1365-2966.2011.18813.x},
  \href {http://adsabs.harvard.edu/abs/2011MNRAS.415.1691B} {415, 1691}

\bibitem[\protect\citeauthoryear{{Balona}, {Joshi}, {Joshi}  \&
  {Sagar}}{{Balona} et~al.}{2013}]{balona2013}
{Balona} L.~A.,  {Joshi} S.,  {Joshi} Y.~C.,   {Sagar} R.,  2013, \mn@doi
  [\mnras] {10.1093/mnras/sts429}, \href
  {http://adsabs.harvard.edu/abs/2013MNRAS.429.1466B} {429, 1466}

\bibitem[\protect\citeauthoryear{{Briquet}, {Korhonen}, {Gonz{\'a}lez},
  {Hubrig}  \& {Hackman}}{{Briquet} et~al.}{2010}]{briquet2010}
{Briquet} M.,  {Korhonen} H.,  {Gonz{\'a}lez} J.~F.,  {Hubrig} S.,   {Hackman}
  T.,  2010, \mn@doi [\aap] {10.1051/0004-6361/200913775}, \href
  {http://adsabs.harvard.edu/abs/2010A%26A...511A..71B} {511, A71}

\bibitem[\protect\citeauthoryear{{Budaj}}{{Budaj}}{1996}]{budaj1996}
{Budaj} J.,  1996, \aap, \href
  {http://adsabs.harvard.edu/abs/1996A%26A...313..523B} {313, 523}

\bibitem[\protect\citeauthoryear{{Castelli}}{{Castelli}}{2005}]{castelli2005}
{Castelli} F.,  2005, Memorie della Societa Astronomica Italiana Supplementi,
  \href {http://adsabs.harvard.edu/abs/2005MSAIS...8...25C} {8, 25}

\bibitem[\protect\citeauthoryear{{Castelli} \& {Hubrig}}{{Castelli} \&
  {Hubrig}}{2004}]{castelli2004}
{Castelli} F.,  {Hubrig} S.,  2004, \mn@doi [\aap]
  {10.1051/0004-6361:20041011}, \href
  {http://adsabs.harvard.edu/abs/2004A%26A...425..263C} {425, 263}

\bibitem[\protect\citeauthoryear{{Catanzaro}, {Giarrusso}, {Leone}, {Munari},
  {Scalia}, {Sparacello}  \& {Scuderi}}{{Catanzaro}
  et~al.}{2016}]{catanzaro2016}
{Catanzaro} G.,  {Giarrusso} M.,  {Leone} F.,  {Munari} M.,  {Scalia} C.,
  {Sparacello} E.,   {Scuderi} S.,  2016, \mn@doi [\mnras]
  {10.1093/mnras/stw923}, \href
  {http://adsabs.harvard.edu/abs/2016MNRAS.460.1999C} {460, 1999}

\bibitem[\protect\citeauthoryear{{Ghazaryan}, {Alecian}  \&
  {Harutyunian}}{{Ghazaryan} et~al.}{2013}]{ghazaryan2013}
{Ghazaryan} S.,  {Alecian} G.,   {Harutyunian} H.,  2013, \mn@doi [\mnras]
  {10.1093/mnras/stt1339}, \href
  {http://adsabs.harvard.edu/abs/2013MNRAS.435.1852G} {435, 1852}

\bibitem[\protect\citeauthoryear{{Grevesse} \& {Sauval}}{{Grevesse} \&
  {Sauval}}{1998}]{grevesse1998}
{Grevesse} N.,  {Sauval} A.~J.,  1998, \mn@doi [\ssr]
  {10.1023/A:1005161325181}, \href
  {http://adsabs.harvard.edu/abs/1998SSRv...85..161G} {85, 161}

\bibitem[\protect\citeauthoryear{{Hubrig} et~al.,}{{Hubrig}
  et~al.}{2010}]{hubrig2010}
{Hubrig} S.,  et~al., 2010, \mn@doi [\mnras]
  {10.1111/j.1745-3933.2010.00928.x}, \href
  {http://adsabs.harvard.edu/abs/2010MNRAS.408L..61H} {408, L61}

\bibitem[\protect\citeauthoryear{{Hubrig} et~al.,}{{Hubrig}
  et~al.}{2012}]{hubrig2012}
{Hubrig} S.,  et~al., 2012, \mn@doi [\aap] {10.1051/0004-6361/201219778}, \href
  {http://adsabs.harvard.edu/abs/2012A%26A...547A..90H} {547, A90}

\bibitem[\protect\citeauthoryear{{H{\"u}mmerich}, {Niemczura}, {Walczak},
  {Paunzen}, {Bernhard}, {Murphy}  \& {Drobek}}{{H{\"u}mmerich}
  et~al.}{2018}]{hummerich2018}
{H{\"u}mmerich} S.,  {Niemczura} E.,  {Walczak} P.,  {Paunzen} E.,  {Bernhard}
  K.,  {Murphy} S.~J.,   {Drobek} D.,  2018, \mn@doi [\mnras]
  {10.1093/mnras/stx2974}, \href
  {http://adsabs.harvard.edu/abs/2018MNRAS.474.2467H} {474, 2467}

\bibitem[\protect\citeauthoryear{{Kochukhov}, {Adelman}, {Gulliver}  \&
  {Piskunov}}{{Kochukhov} et~al.}{2007}]{kochukhov2007}
{Kochukhov} O.,  {Adelman} S.~J.,  {Gulliver} A.~F.,   {Piskunov} N.,  2007,
  \mn@doi [Nature Physics] {10.1038/nphys648}, \href
  {http://adsabs.harvard.edu/abs/2007NatPh...3..526K} {3, 526}

\bibitem[\protect\citeauthoryear{{Kochukhov}, {Wade}  \& {Shulyak}}{{Kochukhov}
  et~al.}{2012}]{kochukhov2012}
{Kochukhov} O.,  {Wade} G.~A.,   {Shulyak} D.,  2012, \mn@doi [\mnras]
  {10.1111/j.1365-2966.2012.20526.x}, \href
  {http://adsabs.harvard.edu/abs/2012MNRAS.421.3004K} {421, 3004}

\bibitem[\protect\citeauthoryear{{Kochukhov} et~al.,}{{Kochukhov}
  et~al.}{2013}]{kochukhov2013}
{Kochukhov} O.,  et~al., 2013, \mn@doi [\aap] {10.1051/0004-6361/201321467},
  \href {http://adsabs.harvard.edu/abs/2013A%26A...554A..61K} {554, A61}

\bibitem[\protect\citeauthoryear{{Korhonen} et~al.,}{{Korhonen}
  et~al.}{2013}]{korhonen2013}
{Korhonen} H.,  et~al., 2013, \mn@doi [\aap] {10.1051/0004-6361/201220951},
  \href {http://adsabs.harvard.edu/abs/2013A%26A...553A..27K} {553, A27}

\bibitem[\protect\citeauthoryear{{Krti{\v c}ka}, {Mikul{\'a}{\v s}ek}, {Henry},
  {Zverko}, {{\v Z}i{\v z}ovsk{\'y}}, {Skalick{\'y}}  \& {Zv{\v e}{\v
  r}ina}}{{Krti{\v c}ka} et~al.}{2009}]{krticka2009}
{Krti{\v c}ka} J.,  {Mikul{\'a}{\v s}ek} Z.,  {Henry} G.~W.,  {Zverko} J.,
  {{\v Z}i{\v z}ovsk{\'y}} J.,  {Skalick{\'y}} J.,   {Zv{\v e}{\v r}ina} P.,
  2009, \mn@doi [\aap] {10.1051/0004-6361/200811123}, \href
  {http://adsabs.harvard.edu/abs/2009A%26A...499..567K} {499, 567}

\bibitem[\protect\citeauthoryear{{Krti{\v c}ka}, {Mikul{\'a}{\v s}ek},
  {L{\"u}ftinger}, {Shulyak}, {Zverko}, {{\v Z}i{\v z}{\v n}ovsk{\'y}}  \&
  {Sokolov}}{{Krti{\v c}ka} et~al.}{2012}]{krticka2012}
{Krti{\v c}ka} J.,  {Mikul{\'a}{\v s}ek} Z.,  {L{\"u}ftinger} T.,  {Shulyak}
  D.,  {Zverko} J.,  {{\v Z}i{\v z}{\v n}ovsk{\'y}} J.,   {Sokolov} N.~A.,
  2012, \mn@doi [\aap] {10.1051/0004-6361/201117490}, \href
  {http://adsabs.harvard.edu/abs/2012A%26A...537A..14K} {537, A14}

\bibitem[\protect\citeauthoryear{{Kurucz}}{{Kurucz}}{1996}]{kurucz1996}
{Kurucz} R.~L.,  1996, in {Adelman} S.~J.,  {Kupka} F.,   {Weiss} W.~W.,  eds,
  Astronomical Society of the Pacific Conference Series Vol. 108, M.A.S.S.,
  Model Atmospheres and Spectrum Synthesis. p.~160

\bibitem[\protect\citeauthoryear{{Lanz} \& {Hubeny}}{{Lanz} \&
  {Hubeny}}{2007}]{lanz2007}
{Lanz} T.,  {Hubeny} I.,  2007, \mn@doi [\apjs] {10.1086/511270}, \href
  {http://adsabs.harvard.edu/abs/2007ApJS..169...83L} {169, 83}

\bibitem[\protect\citeauthoryear{{L{\"u}ftinger}, {Kochukhov}, {Ryabchikova},
  {Piskunov}, {Weiss}  \& {Ilyin}}{{L{\"u}ftinger}
  et~al.}{2010}]{luftinger2010}
{L{\"u}ftinger} T.,  {Kochukhov} O.,  {Ryabchikova} T.,  {Piskunov} N.,
  {Weiss} W.~W.,   {Ilyin} I.,  2010, \mn@doi [\aap]
  {10.1051/0004-6361/200811545}, \href
  {http://adsabs.harvard.edu/abs/2010A%26A...509A..71L} {509, A71}

\bibitem[\protect\citeauthoryear{{Makaganiuk} et~al.,}{{Makaganiuk}
  et~al.}{2011}]{makaganiuk2011}
{Makaganiuk} V.,  et~al., 2011, \mn@doi [\aap] {10.1051/0004-6361/201016302},
  \href {http://adsabs.harvard.edu/abs/2011A%26A...529A.160M} {529, A160}

\bibitem[\protect\citeauthoryear{{Makaganiuk} et~al.,}{{Makaganiuk}
  et~al.}{2012}]{makaganiuk2012}
{Makaganiuk} V.,  et~al., 2012, \mn@doi [\aap] {10.1051/0004-6361/201118167},
  \href {https://ui.adsabs.harvard.edu/abs/2012A&A...539A.142M} {539, A142}

\bibitem[\protect\citeauthoryear{{Michaud}, {Charland}, {Vauclair}  \&
  {Vauclair}}{{Michaud} et~al.}{1976}]{michaud1976}
{Michaud} G.,  {Charland} Y.,  {Vauclair} S.,   {Vauclair} G.,  1976, \mn@doi
  [\apj] {10.1086/154848}, \href
  {http://adsabs.harvard.edu/abs/1976ApJ...210..447M} {210, 447}

\bibitem[\protect\citeauthoryear{{Monier}, {Gebran}  \& {Royer}}{{Monier}
  et~al.}{2015}]{monier2015}
{Monier} R.,  {Gebran} M.,   {Royer} F.,  2015, \mn@doi [\aap]
  {10.1051/0004-6361/201526106}, \href
  {http://adsabs.harvard.edu/abs/2015A%26A...577A..96M} {577, A96}

\bibitem[\protect\citeauthoryear{{Morel} et~al.,}{{Morel}
  et~al.}{2014}]{morel2014}
{Morel} T.,  et~al., 2014, \mn@doi [\aap] {10.1051/0004-6361/201322289}, \href
  {http://adsabs.harvard.edu/abs/2014A%26A...561A..35M} {561, A35}

\bibitem[\protect\citeauthoryear{{Paunzen} et~al.,}{{Paunzen}
  et~al.}{2018}]{paunzen2018}
{Paunzen} E.,  et~al., 2018, \mn@doi [\aap] {10.1051/0004-6361/201732257},
  \href {http://adsabs.harvard.edu/abs/2018A%26A...615A..36P} {615, A36}

\bibitem[\protect\citeauthoryear{{Piskunov} \& {Kochukhov}}{{Piskunov} \&
  {Kochukhov}}{2002}]{piskunov2002}
{Piskunov} N.,  {Kochukhov} O.,  2002, \mn@doi [\aap]
  {10.1051/0004-6361:20011517}, \href
  {http://adsabs.harvard.edu/abs/2002A%26A...381..736P} {381, 736}

\bibitem[\protect\citeauthoryear{{Pourbaix}, {Boffin}, {Chini}  \&
  {Dembsky}}{{Pourbaix} et~al.}{2013}]{pourbaix2013}
{Pourbaix} D.,  {Boffin} H.~M.~J.,  {Chini} R.,   {Dembsky} T.,  2013, \mn@doi
  [\aap] {10.1051/0004-6361/201321699}, \href
  {http://adsabs.harvard.edu/abs/2013A%26A...556A..45P} {556, A45}

\bibitem[\protect\citeauthoryear{{Prv{\'a}k}, {Li{\v s}ka}, {Krti{\v c}ka},
  {Mikul{\'a}{\v s}ek}  \& {L{\"u}ftinger}}{{Prv{\'a}k}
  et~al.}{2015}]{prvak2015}
{Prv{\'a}k} M.,  {Li{\v s}ka} J.,  {Krti{\v c}ka} J.,  {Mikul{\'a}{\v s}ek} Z.,
    {L{\"u}ftinger} T.,  2015, \mn@doi [\aap] {10.1051/0004-6361/201526647},
  \href {http://adsabs.harvard.edu/abs/2015A%26A...584A..17P} {584, A17}

\bibitem[\protect\citeauthoryear{{Prv{\'a}k}, {Krti{\v c}ka}  \&
  {Korhonen}}{{Prv{\'a}k} et~al.}{2018}]{prvak2018}
{Prv{\'a}k} M.,  {Krti{\v c}ka} J.,   {Korhonen} H.,  2018, Contributions of
  the Astronomical Observatory Skalnate Pleso, \href
  {http://adsabs.harvard.edu/abs/2018CoSka..48...93P} {48, 93}

\bibitem[\protect\citeauthoryear{{Rice}, {Wehlau}  \& {Khokhlova}}{{Rice}
  et~al.}{1989}]{rice1989}
{Rice} J.~B.,  {Wehlau} W.~H.,   {Khokhlova} V.~L.,  1989, \aap, \href
  {http://adsabs.harvard.edu/abs/1989A%26A...208..179R} {208, 179}

\bibitem[\protect\citeauthoryear{{Ricker} et~al.,}{{Ricker}
  et~al.}{2015}]{ricker2015}
{Ricker} G.~R.,  et~al., 2015, \mn@doi [{J. of Astron. Telescopes, Instruments,
  and Systems}] {10.1117/1.JATIS.1.1.014003}, \href
  {http://adsabs.harvard.edu/abs/2015JATIS...1a4003R} {1, 014003}

\bibitem[\protect\citeauthoryear{{Riley}}{{Riley}}{2018}]{stis}
{Riley} A.,  2018, {STIS Instrument Handbook}

\bibitem[\protect\citeauthoryear{{Sch{\"o}ller}, {Correia}, {Hubrig}  \&
  {Ageorges}}{{Sch{\"o}ller} et~al.}{2010}]{scholler2010A}
{Sch{\"o}ller} M.,  {Correia} S.,  {Hubrig} S.,   {Ageorges} N.,  2010, \mn@doi
  [\aap] {10.1051/0004-6361/201014246}, \href
  {http://adsabs.harvard.edu/abs/2010A%26A...522A..85S} {522, A85}

\bibitem[\protect\citeauthoryear{{Sikora} et~al.,}{{Sikora}
  et~al.}{2019}]{sikora2019}
{Sikora} J.,  et~al., 2019, \mn@doi [Monthly Notices of the Royal Astronomical
  Society] {10.1093/mnras/stz1581}, \href
  {https://ui.adsabs.harvard.edu/abs/2019MNRAS.487.4695S} {487, 4695}

\bibitem[\protect\citeauthoryear{{Stift}, {Leone}  \& {Cowley}}{{Stift}
  et~al.}{2012}]{stift2012}
{Stift} M.~J.,  {Leone} F.,   {Cowley} C.~R.,  2012, \mn@doi [\mnras]
  {10.1111/j.1365-2966.2011.19933.x}, \href
  {http://adsabs.harvard.edu/abs/2012MNRAS.419.2912S} {419, 2912}

\bibitem[\protect\citeauthoryear{{Strassmeier}, {Granzer}, {Mallonn}, {Weber}
  \& {Weingrill}}{{Strassmeier} et~al.}{2017}]{strassmeier2017}
{Strassmeier} K.~G.,  {Granzer} T.,  {Mallonn} M.,  {Weber} M.,   {Weingrill}
  J.,  2017, \mn@doi [\aap] {10.1051/0004-6361/201629150}, \href
  {http://adsabs.harvard.edu/abs/2017A%26A...597A..55S} {597, A55}

\makeatother
\end{thebibliography}



\bsp	
\label{lastpage}
\end{document}